\documentclass[]{aastex}
\usepackage{graphicx}
\usepackage{amsmath} 
\usepackage{natbib}

\usepackage{emulateapj5}
\usepackage{onecolfloat}

\newcommand{\beqa}{\begin{eqnarray}}
\newcommand{\eeqa}{\end{eqnarray}}
\newcommand{\Beq}{\begin{equation}}
\newcommand{\Eeq}{\end{equation}}
\newcommand{\nn}{\nonumber}

\newcommand{\vu}{{\boldsymbol{u}}}
\newcommand{\vk}{{\boldsymbol{k}}}
\newcommand{\vx}{{\boldsymbol{x}}}

\renewcommand{\vr}{{\boldsymbol{r}}}

\newcommand{\vv}{{\boldsymbol{v}}}

\newcommand{\ve}{{\boldsymbol{\eta}}}
\newcommand{\vnab}{{\boldsymbol{\nabla}}}

\newcommand{\calE}{{\cal E}}
\newcommand{\calF}{{\cal F}}

\newcommand{\deu}{\partial}

\newcommand{\Mean}[1]{ \left<\,  #1 \,\right>}

\newcommand{\Eerg}{\:\mathrm{erg}}
\newcommand{\Eg}{\:\mathrm{g}}
\newcommand{\Ekm}{\:\mathrm{km}}
\newcommand{\EMpc}{\:\mathrm{Mpc}}

\newcommand{\Epers}{\:\mathrm{s}^{-1}}
\newcommand{\Epercs}{\:\mathrm{s}^{-3}}
\newcommand{\Epercm}{\:\mathrm{cm}^{-1}}
\newcommand{\Eperccm}{\:\mathrm{cm}^{-3}}
\newcommand{\EperMpc}{\:\mathrm{Mpc}^{-1}}
\newcommand{\EMyr}{\:\mathrm{Myr}}
\newcommand{\Eyr}{\:\mathrm{yr}}
\newcommand{\Eperyr}{\:\mathrm{yr}^{-1}}

\newcommand{\Msol}{\:\mathrm{M}_\odot}
\newcommand{\Ion}[2]{#1$\,${\sc #2}}

\begin{document}

\title{
	The scale-dependent energy transfer rate 
	as a tracer for star formation 
	in cosmological n-body simulations }

\shorttitle{Energy exchange as tracer for star formation}
\shortauthors{Hoeft et al.}

\author{M. Hoeft\altaffilmark{3}}
\affil{Astrophysikalisches Institut Potsdam\altaffilmark{1}}
\affil{Technische Universit\"at Berlin\altaffilmark{2}}
\author{J.P. M\"ucket\altaffilmark{4}}
\affil{Astrophysikalisches Institut Potsdam\altaffilmark{1}}
\author{P. Heide\altaffilmark{5}}
\affil{Technische Universit\"at Berlin\altaffilmark{2}}

\altaffiltext{1}{An der Sternwarte 16, D-14482  Potsdam, Germany }
\altaffiltext{2}{Institut f\"ur Atomare Physik und Fachdidaktik, PN 3-1\\
Hardenbergstra{\ss}e 36, D-10623 Berlin, Germany }
\altaffiltext{3}{mhoeft@aip.de}
\altaffiltext{4}{jpmuecket@aip.de}
\altaffiltext{5}{heide@physik.tu-berlin.de}

\begin{abstract}

We investigate the energy release due to the large-scale structure formation and 
the subsequent transfer of energy from larger to smaller scales.  
We  calculate the power spectra for the large-scale velocity field and show that 
the coupling of modes results in a transfer of power predominately from larger 
to smaller scales.  
We use the concept of cumulative energy for calculating which energy amount is 
deposited into the small scales during the cosmological structure evolution.  
To estimate the contribution due to the gravitational interaction only we 
perform our investigations by means of dark matter simulations.  
The global mean of the energy transfer increases with redshift $\sim (z+1)^{3}$; 
this can be traced back to the similar evolution of the merging rates of dark 
matter halos.  

The global mean energy transfer can be decomposed into its local contributions, 
which allows to determine the energy injection per mass into a local volume.  
The obtained energy injection rates are at least comparable with other energy  
sources driving the interstellar turbulence as, e.g. by the supernova kinetic 
feedback.

On that basis  we make the crude assumption that processes causing this energy 
transfer from large 
to small scales, e.g. the 
merging of halos, may contribute substantially to drive the ISM turbulence which 
may eventually result in star formation on much smaller scales. We
propose that the ratio of the local energy injection rate to the energy already 
stored within small-scale motions is a rough measure for the probability of 
the local star formation efficiency applicable within cosmological large-scale 
n-body simulations.  
\end{abstract}


\twocolumn

\section{Introduction}
\label{introduction}
During the last decade great success could be achieved in our understanding of
the detailed mechanisms for the formation and evolution of cosmic structure.
The role of the underlying cosmological models and parameters have been
investigated by numerous numerical n-body simulations.  
On large scales the calculated distribution of matter is in excellent agreement 
with the observed one, e.g. with the galaxy distribution and the distribution of 
the intergalactic medium (IGM).
The fast enhancement of the available computational power permitted to cover an 
increasing dynamical range and/or to consider additional processes. In 
particular hydrodynamical models have been developed very successfully.

The evolution of cosmic matter can roughly be subdivided in terms of scales:
On spatial scales larger than the Jeans length gravitation dominates and on 
smaller 
scales hydrodynamical processes do. 
On the other hand for the description of the 
IGM, e.g., this distinction is insufficient.
The IGM is strongly influenced by
both, the large scale structure evolution and the feedback of the luminous
matter, in particular by the star formation processes:  Supernova explosions
sweep out the galactic gas enriched by heavy elements into the IGM
changing its chemical composition and thermal state.  Radiation ionizates
the IGM in the environment of the galaxies.

In order to obtain an appropriate description of the physical state of the 
large-scale distributed gas also the amount and the distribution of stars and 
their back-reaction has to be estimated.
However, to link the process of star formation to the large-scale structure 
evolution is by far out of scope.

Incorporating the stellar feed-back in simulations inevitably needs to connect 
the star formation rate to available gas parameters.
Those parameters can be, e.g. the local density and the local gas temperature.
Note, in terms of the 
considered simulations `local' stands for the average over smallest resolved 
scales, usually of the order of 1 - 100 kpc.   
\citet{schmidt:59} found that in the interstellar gas the star formation rate 
$\dot{\rho}_\ast$ (SFR) is 
related to the density $\rho$ by $\dot{\rho}_\ast \propto \rho^n$, where $n$ is 
adopted to be about 1.5 \citep[cf.][and references therein]{kennicutt:98}.
Although those scales are far below the resolution in large-scale simulations, 
this empirical relation is often applied.
Gas which fulfills certain density and temperature criteria is assumed to form 
stars according to the above given Schmidt-law 
\citep[e.g.][]{yepes:97,springel:00,steinmetz:01,nagamine:01,ascasibar:01}.
An alternative approach has been introduced by  \cite{kauffmann:99} linking 
semi-analytic galaxy models to bound dark matter halos.
Applying such prescriptions for star formation, e.g., permits to calculate the 
star formation history and the stellar metallicity distribution in the universe. 
However, the variety of used criteria \citep[cf.][]{kay:01} indicates that the 
conjunction is still uncertain.

Knowing the processes on the scale of Molecular Clouds (MCs) which most probably 
are 
controlling the star formation rate would provide an indication on the possible 
linking quantities. Therefore, let us shortly summarize some recent results of 
detailed investigations related to star formation. 

Star formation is hosted by interstellar clouds of molecular hydrogen. Stars
probably arise from shock-compressed dense cores within the clouds 
\citep{blitz:99}. The
cores are produced by supersonic motions of the gas due to the presence of
turbulence \citep{burkert:01}.  According to \citet{klessen:00b} the 
SFR especially depends on the scales on which turbulence is
driven.  

There are indications that the
formation of clouds is linked to larger scales:  \cite*{blitz:99} argued that
MCs are formed through the condensation of \Ion{H}{I} regions in conjunction
with some other mechanism as has been proposed by
\cite{ballesteros-paredes:99}, e.g. due to  colliding
interstellar gas streams \citep[cf.][]{burkert:01}.

The life-time of the MCs, $20 - 100\EMyr$,
indicates that the internal turbulence is more or less continuously driven by
external forces \citep{maclow:98,padoan:01}.  Nevertheless, star formation 
occurs
probably only once within one cloud \citep{elmegreen:00}.  Consequently, the
rate of star formation may depend on the presence of an effective
intercloud turbulence.  

The state of the ISM, i.e. whether it is turbulent or not, may be derived from
the power spectrum of the velocity field or from that of the matter 
distribution.
For instance, \citet{goldman:00} investigated the neutral hydrogen distribution
of the Small Magellanic Cloud (SMC) and derived the spatial power spectrum.  He
found that the spectrum of the SMC is to some extent steeper than Kolmogorov's
1/3-law and explained that by the compressibility of the gas.  Furthermore, he
suggested that the origin of the turbulence could be caused by a close encounter 
between the Small
and the Large Magellanic Cloud (LMC) $2\times10^8\Eyr$ ago and he showed that 
supernova
feedback could only contribute little to the kinetic energy of the SMC.
This illustrates one possible mechanism to force the ISM into a turbulent state:
kinetic energy is transferred from larger to smaller scales, i.e.,  kinetic 
energy from the relative motion of the SMC and LMC is
transferred into internal motions.  It is intriguing in respect to the relation
between turbulence and star formation that both the SMC and the LMC show
evidence for large increases of their star formation rate at the time of the
encounter \citep{larson:01}.

From these investigations one may conclude that the SFR is controlled by the 
rate of formation of dense substructures - or clouds - in the interstellar 
medium, dense enough to undergo gravitational collapse and fragmentation. 
Whether these substructures are generated by thermal instability, gravitational 
instability, turbulent compression or some other process is still under debate 
\citep[cf. introduction in][]{scalo:02}.
We suppose that interstellar turbulence is the leading mechanism.
Then the question arises, how is the interstellar medium forced into a turbulent 
state? What drives the turbulence over a sufficient long time period needed to 
form eventually stars?
Again, different energy sources such as supernova feedback, galactic 
differential rotation \citep{sellwood:99}, infall of  high-velocity clouds 
\citep{blitz:99b} and other mechanism are conceivable.
Merging events between galaxies or minor-merging events between galaxies and 
smaller clouds are supposed to cause efficient star formation 
\citep{kolatt:99,somerville:01,kauffmann:01}.  
This may indicate that merging, which occurs due to the structure formation, 
could provide sufficient energy to stir the interstellar medium.

Therefore, the question arises how much energy is eventually available by 
injection from the extra-galactic scales. When large-scale structures are formed 
gravitational energy is released and stored in large-scale motions. The 
structure evolution transfers
the kinetic energy from large-scales to galactic scales. May this 
transferred energy significantly contribute for balancing the
dissipation of the turbulent field on the sub-galactic scales? 

In this paper we want to address the question which energy transfer down to 
galactic scales can be expected from the large-scale structure formation, what 
is the time evolution of the transfer and how is it spatially distributed.
On large scales ($\gtrsim 0.5\EMpc$) the baryonic matter is still tightly 
coupled to the dark matter, thus the energy transfer is governed by the 
gravitational interaction only.   
We use cosmological n-body simulations to determine the transfer.  
We start with the investigation of the power spectrum of the large-scale 
velocity field, closely related to the energy spectrum (Sec.~\ref{sec-vel}).  
Then by introducing the concept of cumulative energy we determine the energy 
injection into galactic-scale motions (Sec.~\ref{sec-culme}).  
We calculate the volume averaged energy transfer and discuss its relation to the 
merging rate evolution of halos (Sec.~\ref{sec-pik}). 
Furthermore, the spatial distribution of the energy injection at given scale is 
determined.  
This allows to infer a local energy injection rate which can be attributed to a 
heating rate with respect to the baryonic matter (Sec.~\ref{sec-dsg}).

Making the crude simplification that the obtained energy injection 
rates at galactic scales are linked to the necessary energy input for driving 
the interstellar turbulence and moreover assuming that this turbulence is 
controlling the star formation leads us to an estimator for the local - in terms 
of cosmological simulations - star formation rate (Sec.~\ref{sec-sfr}).
Finally we summarize our results and discuss possible implications 
(Sec.~\ref{summary}).

\section{Power spectrum of the large-scale velocity field}

\label{sec-vel}

We calculate the power spectrum of the large-scale velocity field using n-body 
simulations of the cosmic structure formation.
The spectrum should possess similar features as the power spectrum of the 
density field:
At scales above $\approx 60\EMpc$ (at present) the modes are decoupled and still 
evolve according to linear growth whereas at smaller scales the evolution of the 
modes is highly non-linear and coupled.

We simulate the formation of the large-scale structure using a 
particle-particle/particle-mesh (P$^3$M) code 
considering only the interaction of the dark matter.
We have performed different simulations with box sizes $L_{\rm{box}}$ from 8 to 
$250\,h^{-1}\EMpc$ and with $256^3$ and $128^3$ particles.
We apply a cosmological model with matter density $\Omega_0 = 0.3$, cosmological 
constant $\Omega_\Lambda = 0.7$, baryon density $\Omega_B = 0.04$ and a Hubble 
constant $H_0 = h H_0^{100} = 70 \Ekm\Epers\EperMpc$.
In this paper we are mainly interested in establishing a method to calculate the 
local energy release due the formation of the large-scale structure.
We therefore apply only the customary $\Lambda$CDM model given above.

Using n-body simulations the velocity field is only defined at particle 
positions.
Therefore, calculating the power spectrum of the velocity field by its Fourier 
transform would require to interpolate the velocity field to a grid.
We avoid this procedure by applying an algorithm which was introduced first by 
\citet{lomb:76} to calculate the `periodogram' of a randomly sampled 
time-dependent signal.
We apply the algorithm in its form given by \citet{press:86} and extend it to 
three-dimensional data sets.

\begin{figure}[t]
	\centering
	\resizebox{\hsize}{!}{\includegraphics{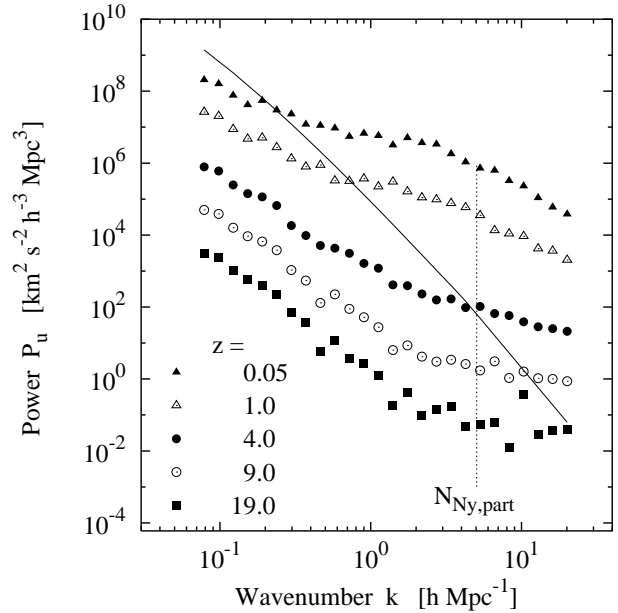}}
	\caption{
	Power spectra of the velocity field at different redshifts $z$
	calculated by the algorithm introduced by \citet{lomb:76}.
	The solid line gives the spectrum according to the linear growth
	at $z = 0.05$
	The vertical dotted line gives the Nyquist wavenumber which corresponds 
	to the minimal distance of two particles 
	if the particles are evenly distributed.
	}
	\label{fig-lomb1}
\end{figure}

The most likely power $P_\vv (k)$ of a mode with wavevector $\vk$ for a velocity 
field sampled at $N$ points $\vx_j \; (j\in N)$ is given by
\beqa
P_\vv (\vk)
   &=&
	\frac{1}{2 N x_0^3}\:
	\left\{
   \frac{\left| \sum\nolimits_j
      (\vv_j - \overline{\vv} ) \cos \Theta_j(\vk) \right|^2 }
   {\sum\nolimits_j \cos^2 \Theta_j(\vk) } 
	\right.
	\nn \\
	&& 
	\qquad
	+
	\left.
   \frac{\left| \sum\nolimits_j
      (\vv_j - \overline{\vv} ) \sin \Theta_j(\vk) \right|^2 }
   {\sum\nolimits_j \sin^2 \Theta_j(\vk) }
	\right\}
	\label{eq-lomb}
	\\
&& \Theta_j(\vk) 
   \; = \;
   \vk\vx_j - \frac{1}{2} \arctan
   \frac{\sum\nolimits_i \sin 2\vk\vx_i}
      {\sum\nolimits_i \cos 2\vk\vx_i}
	,
   \nn
\eeqa
where $x_0^3$ denotes the mean volume per sampling point.
Averaging the power of several arbitrary wavevectors with the same absolute 
value $|k| = \vk$ eliminates the angular dependency
\beqa
P_\vv (k) 
	&=& 
	\Mean{ P_\vv (\vk) }_{|\vk|=k}
	.
	\nn
\eeqa
If the velocity field is evenly sampled the algorithm is identical to the 
calculations of the power spectrum by the Fourier transform.
If the field is unevenly sampled the algorithm is equivalent to a harmonic 
least-squares analysis \citep{scargle:82}.
Therefore, the largest wavenumber, power of which can unambiguously determined, 
largely exceeds the Nyquist wavenumber of an evenly spaced particle 
distribution.
Figure \ref{fig-lomb1} shows velocity spectra of one simulation taken at 
different redshifts $z$.
A Fourier transform based power spectrum always rapidly decreases when 
approaching the Nyquist wavenumber,
whereas the algorithm of Lomb does not.
The P$^3$M simulations resolve velocity modes whose wavelength is smaller than 
the Nyquist one and the algorithm of Lomb allows to calculate its power.

\begin{figure}[t]
	\centering
	\resizebox{\hsize}{!}{\includegraphics{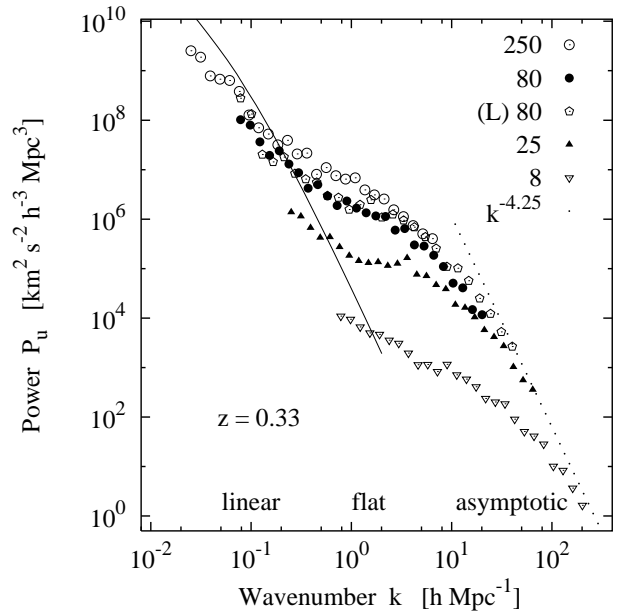}}
	\caption{ 
	Power spectra of the velocity field taken from simulations with 
	different box sizes at same redshift $z=0.33$.
	The labels denote box sizes in $[h^{-1}\EMpc]$. 
	The simulation with label (L) was calculated with $256^3$ particles;
	the other simulations were calculated with $128^3$ particles.
	The solid line shows the power according to the linear growth.
	The dotted line gives an estimated asymptote. 
	}
	\label{fig-lomb2}
\end{figure}

At redshift zero the spectra show at $k \approx 0.2 \, h \EperMpc$ the 
transition from linear to nonlinear growth alike the spectra of the density 
field.
Modes with wavelength above this transition scale evolve independently and 
according to the linear growth function, whereas modes with smaller wavelength 
are coupled to each other and increase much faster.
This becomes more conspicuous by combining the spectra of simulations for 
different box sizes, see Fig.~\ref{fig-lomb2}.
At large scales again the linear growth is reproduced (`linear region'), i.e. 
the modes evolve independly.
Towards smaller scales the slope decreases and a nearly `flat' region occurs.
After the flat region the spectra probably exhibit a common, power law-like 
behavior 
(`asymptotic region').
This deviation from the linear growth in the flat and asymptotic region 
characterizes the coupled mode evolution.  
The power throughout the flat and asymptotic regions depends on the largest 
modes within the simulation box:
Figure~\ref{fig-lomb2} shows that the power of the largest wavelength 
$\lambda_{\rm{max}}$ realized in a simulation evolves always according to the 
linear growth, indicated by the solid line.
The largest wavelengths of the depicted power spectra are located  either in the 
linear or in the flat region.
The two spectra with $\lambda_{\rm{max}}$ in the linear region exhibit roughly 
the same power in the flat region.  
The two spectra with $\lambda_{\rm{max}}$ in the flat region exhibit in contrast 
a much smaller power throughout the flat and asymptotic region; 
the amplitude is governed by the linear growth of the power at the largest 
wavelength $\lambda_{\rm{max}}$.
This shows that in the asymptotic and flat regime the power of a given mode is 
dominated by the larger modes.
There is no evidence that smaller modes are able to increase the power of the 
larger ones, i.e. the transfer of energy is one-directional from larger to 
smaller scales.

In the asymptotic region the shape of the spectra can be described by a power 
law with an index of about -4 to -4.5.
The overlap of the different spectra in this region reflects an asymptotic 
behavior which is independent from the largest modes in the box.
Qualitatively the same three regions can be distinguished in the power spectrum 
of the density field using high resolution simulations \citep[cf. the spectrum 
given by][]{klypin:99}.

The power spectrum $P_\vv (k)$ of the large-scale velocity field can be 
transposed into a velocity-scale relation by 
\beqa
v_l^2
	\propto 
	\int\nolimits_{2\pi/l}^{\infty} dk\, k^2 \, P_\vv (k).
	\label{eq-vl}
\eeqa
The asymptotic region with power-law behavior $\propto k^x$ results in the 
relation $v\propto l^\kappa$ with $\kappa = - (3+x)/2 \approx 0.6 $.

The asymptotic power-law behavior indicates on self-simi\-larity at small 
scales, which should result in a scaling relation, connecting scale $l$, 
velocity $v_l$, and energy transfer $\epsilon$.
We provide a heuristic derivation of a scaling relation based on dimensional 
analysis.

The energy rate per volume element and time has the following unit: $[\epsilon] 
= \Eerg\Eperccm\Epers = \Eg\Epercm\Epercs$. 
The driving force for all processes described by the simulations is gravitation 
and, therefore, the gravitational constant $G$ must enter the expression for the 
energy transfer $\epsilon$. 
Further quantities characterizing the behavior of the n-body system is the 
velocity $v_l$, the associated length scale $l$, and the density $\rho$. 
The only possible combination of these quantities including the gravitational 
constant $G$ to build a quantity with units $[\epsilon]$ is 
\beqa
\epsilon \propto \frac{1}{G}\frac{v_l^5}{l^3}.
\label{eq-dim-vl}
\eeqa
The assumption of a scale independent transfer rate leads to the relation 
between the velocity dispersion and the associated length scale 
$ v_l \propto  l^{3/5} $,
which matches the asymptotic power law obtained by the simulations.
The steeper decline of the scaling-relation, compared to Kolmogorov's relation 
$v_l \propto l^{1/3}$, may be attributed to the gravitational interaction 
leading to a contribution of compressional modes.

The above results support the idea of a nearly constant energy transfer 
throughout the scales of the asymptotic region.
We wish, however, to obtain an exact determination for the energy transfer rate 
through a given scale.
This will be provided in the next section.
It is worth to mention some restriction of Lomb's algorithm if applying to the
power spectrum of the velocity field in n-body simulations:
Lomb's method rests on the assumption that the sampling points are randomly 
distributed, whereas the particles in the simulation are mainly located in the 
dense regions.
Therefore, the true spectrum might slightly differ from the spectrum obtained by 
using Lomb's method.
However, since the velocities are weighted by the local density the obtained 
spectra reflect rather the correct power spectrum for the kinetic energy. 
In this respect they where used and interpreted, indeed.  
A more detailed investigation of the velocity power spectra for different 
initial conditions and cosmological parameters will be given in a forthcoming 
paper.

In the following we use two results of the investigation of the velocity power 
spectrum:
First, the large-scale velocity modes dominate the power of modes at smaller 
scales and,  
second, when forming the large-scale structure the energy transfer is mainly 
directed from larger to smaller scales.

\section{Scale-by-scale energy budget}

\label{sec-culme}

We want to calculate the gain of energy stored in motions with wavelengths below 
a given cut-off $\lambda_K = 2 \pi / K$.
For that purpose we utilize the concept of cumulative energy which was 
introduced by \cite{obukhov:41}, for a review see \cite{frisch:95}.
The cut-off wavenumber $K$ divides the Fourier decomposition of the velocity 
field into a low-pass and a high-pass filtered part, consequently the field can 
be divided into a low-pass and a high-pass filtered field
\beqa
\vv(\vx) &=&
	\frac{1}{(2\pi)^3}
	\int_{|\vk|\leq K} dk \,
	\hat{\vv}(\vk)
	+
	\frac{1}{(2\pi)^3}
	\int_{|\vk|> K} dk \,
	\hat{\vv}(\vk)
	\nn
	\\
	&=&
	\vv(\vx)_K^<
	+
	\vv(\vx)_K^>
	.
	\label{eq-filtered}
\eeqa
If $\vv$ represents the velocity field of an incompressible medium the 
cumulative energy stored in the high-pass filtered velocity modes is given by 
\beqa
\calE_K^{> \rm{(in)}}
	&=&
	\frac{1}{2}
	\rho
	\Mean{ \vv_K^> \cdot \vv_K^> }
	,
	\label{eq-def-culme-in}
\eeqa
where $\Mean{\cdot}$ denotes the volume average.
Cumulative energy $\calE_K^>$ can be gained or lost by energy injection by force 
$\calF_K^>$ or by energy exchange $\Pi_K^>$ between the high-pass and the 
low-pass filtered part:
\beqa
\deu_t \calE_K^{> \rm{(in)}}
	&=&
	\calF_K^>
	- 
	\Pi_K^>
	.
	\label{eq-sbs-bud}
\eeqa
Dark matter does not exhibit any dissipation just by definition but interacts 
via gravitation only.
Thus applying the concept of cumulative energy to the density and velocity field 
of the dark matter the scale-by-scale energy budget Eq.~(\ref{eq-sbs-bud}) 
allows to calculate the energy exchange between different modes and, in 
particular, the energy transfer to the small-scale spectral range.

The energy exchange $\Pi_K^>$ gives an appropriate prescription how to determine 
the energy transfer:
We have shown in the preceding section that the energy stored in motions in the 
`asymptotic' region is governed by the transfer from larger to smaller modes, 
consequently we neglect the energy input by force in the case of the high-pass 
filtered energy $\calE_K^>$.
The high-pass filtered energy is only changed by energy exchange between the 
high-pass and the low-pass filtered part.
To determine the energy exchange $\Pi_K^>$ we must calculate the time derivative 
of the high-pass filtered energy.

Due to the properties of the cosmic matter the above described concept must be 
generalized to the case of a compressible medium.
Thus the definition of the cumulative energy must be changed to allow also for 
the variation of the density field.
Since the volume average $\Mean{\cdot}$ must now be applied to the product of 
three fields $\rho \,\vv\cdot\vv$ different possibilities exist formally to 
construct the filtered energy.
In the case of incompressible matter all possibilities of arranging the 
filtering procedure result in the same cumulative energy $\calE_K^> = \langle 
\vv\cdot\ \vv_K^> \rangle = \langle \vv_K^>\cdot\ \vv_K^> \rangle = \langle 
\vv_K^>\cdot\vv \rangle$.
Determining the cumulative energy of a compressible medium we have to generalize 
the definition Eq.~(\ref{eq-def-culme-in}), e.g. by 
$\langle\rho\,\vv_K^>\cdot\vv_K^>\rangle$ or 
$\langle\rho_K^>\,\vv_K^>\cdot\vv_K^>\rangle$.
We show that the definition
\beqa
\calE_K^>
	&=&
	\Mean{(\rho\vv)_K^>\cdot\vv_K^>}
	\label{eq-def-culmE}
\eeqa
is free of contradictions and we apply this to calculate the energy exchange.

We evidence that the definition Eq.~(\ref{eq-def-culmE}) provides a reasonable 
generalization on the basis of the dynamical equations of the cosmic matter.
We obtain well defined expressions for both the energy injection by force and 
the energy exchange.
This is not possible for the other combinations of filtered velocity and 
density.
The energy injection and exchange can be expressed by the filtered fields 
without time derivatives; to obtain these expressions we start with 
Eq.~(\ref{eq-sbs-bud}), insert the new definition for the cumulative energy, 
eliminate the time derivatives by the dynamical equations, and sort the 
resulting terms.

The dynamics of the dark matter can be described by the Euler equation, i.e. 
the Navier-Stokes equation without viscosity, and the continuity equation.
Introducing comoving coordinates distances are scaled by the Hubble expansion 
$\vr = a\vx$ and peculiar velocities $\vu$ are introduced by $\vv = H\vr + 
a\vu$, where $a$ denotes the scale factor which obeys the Friedman equation and 
which is normalized to unity today. 
Furthermore we introduce the `comoving' density $\rho_c = \rho / a^3$ and 
eventually we get the dynamical equations
\beqa
	\deu_t\vu + (\vu\cdot\vnab)\vu + 
	2H\vu
	&=& 
	-\frac{1}{a^2}\nabla\Phi \\ 
	\label{eq-euler}
	\deu_t\rho_c + \vnab\cdot(\rho_c\vu) 
	&=&
	0
	, 
	\label{eq-conti}
\eeqa
where $\Phi$ denotes the gravitational potential.
Executing the steps given above we obtain the cumulative energy injection by 
force and the energy exchange in terms of the filtered fields
\beqa 
\calF_K^<  
	&=& 
	- \frac{1}{a^2}
	\Mean{(\rho_c\vu_K^<+\ve_K^<)\cdot\nabla\Phi}
	\nn
	\\
\Pi_K^< 
	&=& 
	\Mean{ (\rho_c\vu_K^< + (\rho_c\vu)_K^<)
		\cdot(\vu\cdot\vnab)\vu} +
	\Mean{\vu_K^<\cdot\vu(\vnab\cdot\rho_c\vu)}
	\nn
	.
\eeqa
In addition we introduced the time derivative operator $D_t = \deu_t + 4H$.
The term $4H$ takes into account the time dependence of the peculiar velocities 
due to the Hubble expansions and avoids the occurrence of a fictitious energy 
exchange.
Thus, we rewrite the scale-by-scale energy budget in the case of the cosmic 
matter
\beqa
D_t \calE_K^<
	&=&
	\calF_K^<
	- 
	\Pi_K^<
	;
	\label{eq-sbs-bud-com}
\eeqa
the individual terms are given above. 

We specify the demand for a `well defined energy injection and exchange' by 
giving some requirements which have to be fulfilled.
\begin{enumerate}
	\item[i.]
	Additivity of the cumulative energy
	$
		\calE = \calE_K^< + \calE_K^>
	$
	\item[ii.]
	Additivity of the cumulative injected energy by force 
	$
		\calF =\calF_K^< + \calF_K^>
	$
	\item[iii.]
	For the energy flux must be valid
	$
		\Pi_K^< + \Pi_K^> = 0
	$
	\item[iv.]
	Assuming $\rho=\mathrm{const.}$ the relations must go over into 
	the incompressible case
\end{enumerate}
Using the definition of the cumulative energy Eq.~(\ref{eq-def-culmE}) and the 
above derived expressions for the energy injection and exchange it can 
straightforwardly be proven that the requirements i. to iv. are fulfilled.
Defining in contrast the cumulative energy by another combination of filtered 
fields would not fulfill these relations.

We proceed with calculating the energy exchange using the high-pass filtered 
scale-by-scale energy budget Eq.~(\ref{eq-sbs-bud-com}).
As discussed above the high-pass filtered energy injection by force $\calF_K^>$ 
can be neglected, thus we can calculate the energy exchange by
\beqa
\Pi_K^<
	=
  	- 
	\Pi_K^>
	=
	D_t
	\calE_K^>
	=
	\frac{1}{2}
	\,
	D_t
	\,
	\Mean{(\rho\vu)_K^>\cdot\vu_K^>}
	.
	\label{eq-ex-high}
\eeqa
This allows to determine at each moment the gain of energy stored in small-scale 
motions when simulating the formation of the large-scale structure.
We distinguish between the energy exchange $\Pi_K^<$ and the energy transfer 
rate $\epsilon$ introduced above.
The latter is usually introduced when treating an isotropic, quasi-stationary 
process of energy transport and is the appropriate quantity if assuming a 
scale-independent energy transfer which allows to derive the velocity-scale 
relation.

Contrary to turbulent baryonic matter, which possesses viscosity leading to 
dissipation at some scales, the behavior of the dark matter during the structure 
formation is completely different due to the absence of any dissipation.
Once the energy is transferred to small scales it will not be dissipated but 
remains preserved as kinetic energy.
Figure~\ref{fig-lomb1} shows the resulting continuous increase of power in the 
velocity field.
The energy exchange depends via the cumulative energy $\calE_K^>$ on the cut-off 
wavelength $\lambda_K$, i.e. on the considered scale-length.

As shown in the previous section the energy exchange is directed from large to 
small scales. 
Thus the energy exchange expressed by the cumulative energy delivers the energy 
gain of small-scale motions during the structure formation process.

\begin{figure}[t]
	\centering
	\resizebox{\hsize}{!}{\includegraphics{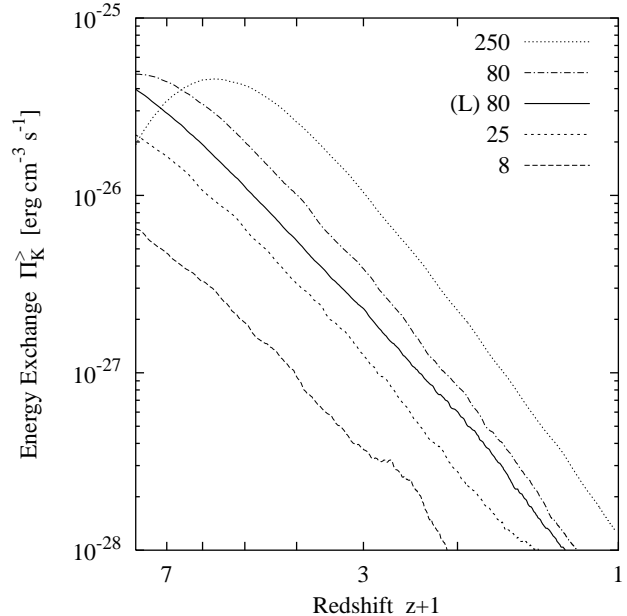}}
	\caption{ 
	The evolution of the global energy exchange obtained from several 
	simulations with different box parameters 
	(box size and resolution, see caption Fig.~\ref{fig-lomb2}) is shown.
	Due to the construction of the cumulative energy 
	(cf. Eqs.~(\ref{eq-filtered}) and (\ref{eq-def-culmE})) 
	the computed quantities depend on the cut-off wavelength $\lambda_K$.
	The latter depends in turn on the realized box size, since the mean 
particle distance    is chosen as $\lambda_K$. 
	Throughout the asymptotic region the velocity power spectrum behaves 
	like $P_\vu (k) \sim  k^{-4.25}$ which 
	leads to the velocity-scale relation $v_l \sim l^{3/5}$, (cf. 
Eq~(\ref{eq-vl})).
	Since the high-pass filtered cumulative energy is roughly proportional 
to
	$v_l^2$ the overall amplitude of the energy exchange scales 
approximately as
	$l^{6/5}$. 
	This dependency can clearly be noticed within the shown figure.
	\label{fig-pik}
	}
\end{figure}

\section{Energy exchange in n-body simulations}

\label{sec-e-ex}

To determine the cumulative energy $\calE_K^>$ in a n-body simulation, which 
deals with a finite number of the particles, we rewrite the definition 
Eq.~(\ref{eq-def-culmE}) in a discrete form.
Each particle samples at its position $\vx_n$ the underlying fields, namely the 
velocity field $\vu(\vx_n)$, the density of momentum $\rho \vu(\vx_n)$, etc..
We take into account that the fraction $V_n$ of the integration volume, which is 
assigned to the particle $n$, depends on its distances to neighbored particles.
We transpose the volume average $\Mean{\cdot}$ into a particle average
\beqa
\calE_K^>
	&=&
	\frac{1}{2}
	\frac{1}{V}
	\sum\nolimits_n 
	(\rho \vu)_K^> (\vx_n) \cdot \vu_K^> (\vx_n)
	\,
	V_n
	.
	\label{eq-discr}
\eeqa	
We estimate the volume $V_n$ by assuming an equidistant grid and dividing the 
cell volume $V_{\rm{cell}}$ by the number of particles $n_i$ per cell i: $V_n = 
V_{\rm{cell}}/n_i$.

Using the additivity of the filtered fields, cf. Eq.~(\ref{eq-filtered}), the 
high-pass filtered velocity field is given by $\vu_K^> = \vu - \vu_K^<$.
The low-pass filtered field is obtained by calculating the average velocity 
within each cell $\vu_K^<(\vx_i) = \langle \vu \rangle_{\rm{cell}}$,
likewise the density of momentum.
The average field at the knots of the grid can only reasonably be determined if 
at least a few particles are within a given cell. 
Determining the low-pass filtered field therefore connotes to introduce a 
density threshold.
We use throughout this paper a low threshold, namely $n_i \geq 3$.
In contrast to the velocity field the density is only defined at the grid, 
therefore, we can rewrite the low-pass filtered density of momentum $(\rho 
\vu)_K^<(\vx_i) = \rho_0 \, n_i \, \langle \vu \rangle_{\rm{cell}}(\vx_i)$, 
where $\rho_0$ denotes the background density.
Interpolating the filtered velocity field $\langle \vu \rangle_{\rm{cell}}$ to 
the positions of the particles we can assign an `individual' dispersion to each 
particle 
\beqa
\sigma_n^2 
	&=&
	\{\vu(\vx_n) - \langle \vu \rangle_{\rm{cell}} (\vx_n) \}^2
	.
\eeqa
Inserting the definition of the individual volume, rewriting the filtered 
density of momentum, and using the dispersion per particle the definition of the 
discrete cumulative energy  Eq.~(\ref{eq-discr}) leads to
\beqa
\calE_K^>
	&=&
	\frac{1}{2}
	\frac{1}{N}
	\rho_0
	\sum\nolimits_n
	\sigma_n^2
	,
\eeqa
where $N$ denotes the number of cells.
Thus, in n-body simulations the cumulative energy is easily obtained from the 
individual dispersions of the particles.
We calculate the energy exchange, using Eq.~(\ref{eq-ex-high}) and the discrete 
form of the cumulative energy, when running the simulations introduced in 
Sec.~\ref{sec-vel}.

\section{Energy exchange and the evolution of the merging rate}

\label{sec-pik}

The energy exchange $\Pi_K^>$ shows a continuous steep decline with time, see 
Fig.~\ref{fig-pik}.
The steep increase at the beginning of the $250\,h^{-1}\,\EMpc$ simulation is 
caused by the necessary density threshold to calculate the filtered fields.
Therefore we do not discuss further this rather artificial effect due to the 
finite mass resolution.
Though the decrease according to a power law with the exponent 3 looks like an 
expansion effect, this is not the case.
By construction, the energy exchange contains only the transfer between 
different modes.
Consequently Fig.~\ref{fig-pik} shows that the mean overall rate of energy 
exchange decreases with time.

The energy exchange depends on the box size of the simulation, too, see 
Fig.~\ref{fig-pik}.
This is due to the fact that we choose the `particle grid size' as cut-off 
wavelength.
Increasing the box size entails that modes, belonging formerly to the low-pass 
filtered modes, become now part of the high-pass filtered ones. The differences 
in amplitudes can easily be estimated by using the asymptote at high wave 
numbers
obtained in Sec.\ref{sec-vel} and performing integration with the corresponding 
cut-off length. 
The energy exchange from larger scales to this modes is additionally considered 
and increases the total amount of energy exchange.

To interpret the decreasing energy exchange we consider the common picture of 
hierarchical structure formation in more detail.
At high redshifts small dark matter halos are formed the mass of which is 
continuously increasing by accretion of `free' matter and merging with other 
halos.
Both processes lead to the transfer of energy formerly stored in large-scale 
motions into internal halo motions.
For the time being we assume that the mean energy exchange per accretion and 
merging process is constant with time.
Then the energy exchange rate reflects the number of accretion and merging 
events per time. 
Assuming furthermore that the merger events are the dominant energy transfer 
processes, the evolution of the energy exchange ought to be equivalent to the 
merging history.
The observed merging rates \citep{carlberg:94} and the rates determined in 
simulations \citep{gottloeber:00} show indeed a behavior of $\sim (z+1)^{3}$.
Thus the decrease of the mean energy exchange with time may be mainly attributed 
to the fact that merging of halos becomes more and more rare during the late 
structure formation.
In the next section we will give further evidence that the assumption of 
approximately constant mean energy exchange per accretion and merging event with 
time is reasonable.

It is worth mentioning that in contrast to investigations elsewhere we are not 
forced to introduce a definition for the individual objects.
The energy exchange traces the merging of halos including a wide spectrum of 
object masses without the difficulty of identifying the individual objects and 
distinguishing between mergers and minor-mergers.

\begin{figure}[t]
	\centering
	\resizebox{\hsize}{!}{\includegraphics{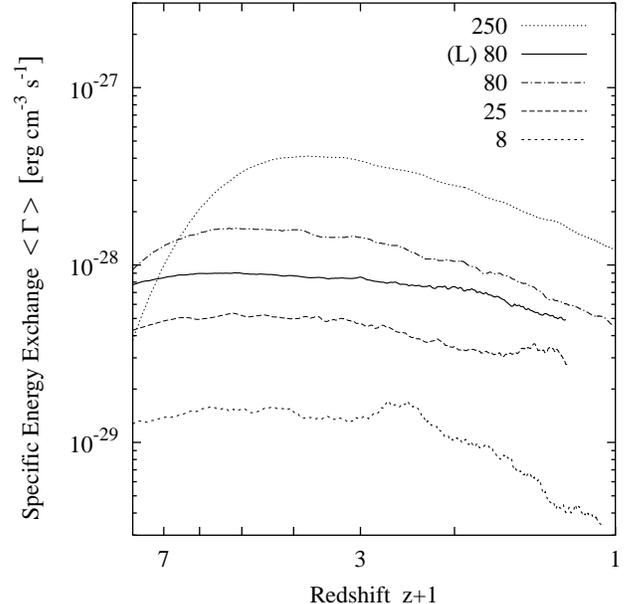}}
	\caption{ 
	Volume averaged specific energy exchange taken from simulations with 
	different box sizes (see caption Fig.~\ref{fig-lomb2}).
	\label{fig-spec}
	}
\end{figure}

\section{Specific energy exchange}

\label{sec-dsg}

\begin{figure*}
	\begin{center}
	\resizebox{0.75\hsize}{!}{\includegraphics{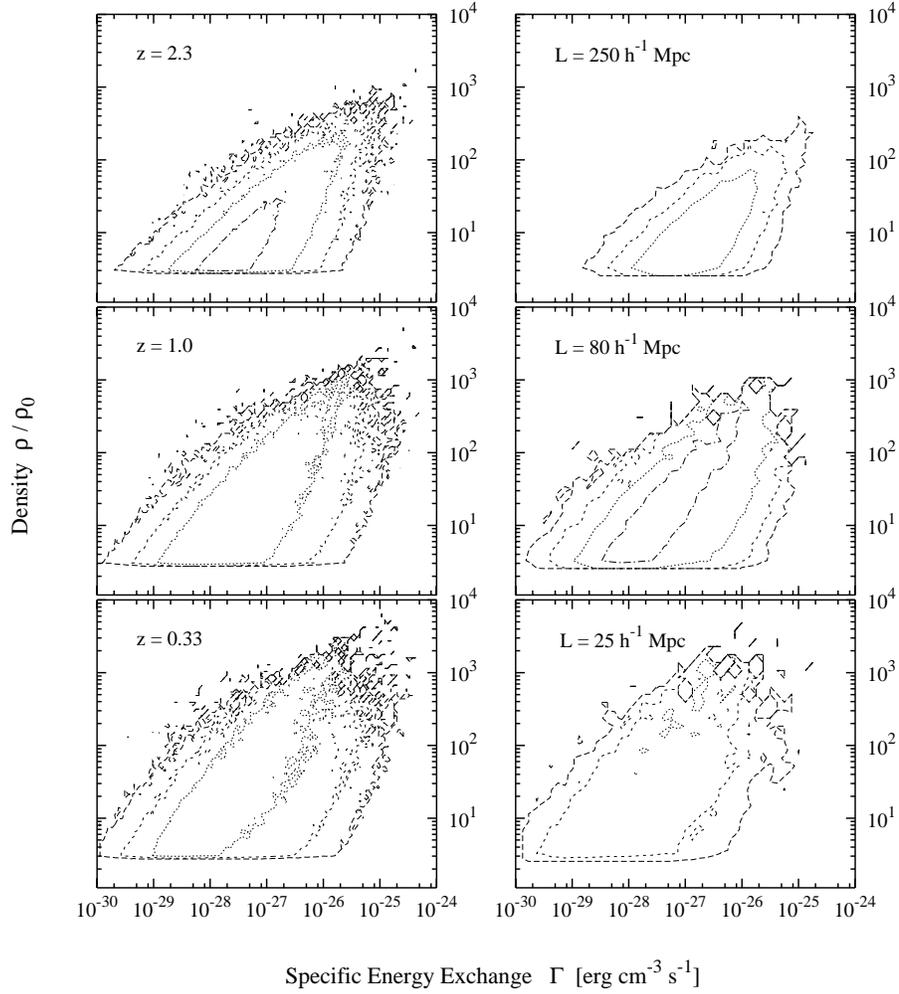}}
	\end{center}
	\caption{ 
	Contours in density contrast $\rho / \rho_0$ and 
	specific energy exchange.
	Left panels are taken from the $80\,h^{-1}\EMpc$ simulation 
	at different redshifts.
	Right panels are taken from three different simulations at the same
  	redshift $z=0.33$.
	\label{fig-spec-2dh}
	}
\end{figure*}

The expression for the global cumulative energy can be represented as the sum of 
local contributions: 
By rewriting its discrete definition Eq.~(\ref{eq-discr}) the sum over all 
particles can be decomposed into the sum over all cell contributions each of 
which collects the contribution of all particles within a given cell volume. 
That is possible because the cumulative energy was constructed by the high pass 
filtered quantities, i.e. by the local velocity dispersions.
Then a {\em local} high-pass filtered energy density can be defined as
\beqa
e_K^>(\vx_i)
	&=&
	\frac{1}{2}
	\rho_0
	\sum\nolimits_{n\in i}
	\sigma_n^2
\eeqa
where $\vx_i$ denotes the position of the cell $i$ and the sum is performed over 
all particles within the cell $n\in i$.
The contributions represent the energy content stored in high-pass filtered 
motions with respect to the local volume related to the cut-off length.
From the local energy density we can define the local contribution of energy 
exchange
\beqa
\pi_K^>
	&=&
	\frac{1}{2}
	\left\{
	D_t + (\vu\cdot\vnab)
	\right\}
	e_K^>.
\eeqa
An example for the local energy transfer is the merging process of halos:
When the separated halos move towards each other both of them carry kinetic 
energy attributed to large-scale motion.
During the merging process at least part of this energy is transferred to the 
internal energy of the resulting halo.

We introduce now the specific energy exchange $\Gamma$ by dividing the local 
energy exchange $\pi_K^>(\vx_i)$ by the amount of mass within the volume of cell 
$i$.
The specific energy exchange could be formally written as a heating rate with 
respect 
to the baryonic mass fraction. 
\beqa
\Gamma (\vx_i)
	&=&
	\frac{\pi_K^> (\vx_i) }
	{n_{\rm{H}} (\vx_i) }
\eeqa
where $n_{\rm{H}}$ denotes the number density of hydrogen atoms.
Of course, this `heating rate' cannot be directly interpreted as a true heating 
of the involved interstellar gas but serves solely as a convenient quantity for 
further estimates and comparisons. 
 
Figure~\ref{fig-spec} shows the evolution of mean the specific energy exchange.
This quantity depends also on the box size of the simulation for reasons which
have been already discussed hitherto.  The $8\,h^{-1}\EMpc$ simulation is
missing power on large-scale modes, whereas the $250\,h^{-1}\EMpc$ does not
resolve small halos.  Therefore the remaining three simulations provide the best
estimate of the specific energy exchange.  The evolution of the mean specific
energy exchange $\Gamma$ varies only slightly with time, compared to the
behavior of the energy exchange itself.  This supports the assumption that the
typical energy exchange per merger is roughly constant during the evolution of
the universe and that the evolution of both the mean energy exchange and the
merging rate are equivalent.

\begin{figure}[t]
	\centering
	\resizebox{\hsize}{!}{\includegraphics{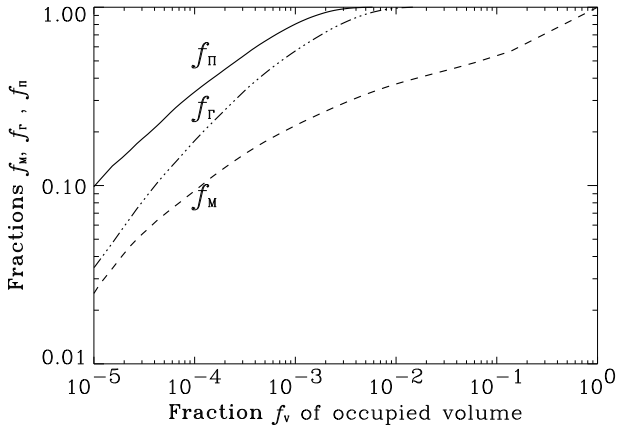}}
	\caption{
	Fraction of mean energy exchange $f_\Pi$, 
	of the mean specific energy exchange $f_\Gamma$, 
	and of the total mass $f_M$ as a function of the occupied fraction of 
volume.	
	\label{fig-concen}
	}
\end{figure}

Basically, the energy exchange only happens when halos merge or matter is
accreted.  Therefore, a correlation is expected between energy exchange and
density.  Figure~\ref{fig-spec-2dh} shows that the specific energy exchange
increases with density; the expectation value of the specific energy exchange is
roughly proportional to the density.  The energy exchange is concentrated to a
small fraction of the total volume; it is much more concentrated than the
matter, see Fig.~\ref{fig-concen}.  Since we consider the energy exchange per
mass the description is likewise valid for the baryonic matter as long it
evolves coupled to the dark matter.  This is also true for the motion of halos
because the baryons are trapped in their gravitational wells.  Thus the
occurrence of a merging event and the resulting energy release can be well
estimated by considering solely the dark matter.

In contrast to the dark matter the baryonic one does not store all energy within 
small-scale motions.
Instead, part of the energy injection at a given scale can be released as heat.
The efficiency of that heating depends on the hydrodynamical processes within 
the baryonic matter, i.e. mainly on the collision rate between particles.
High density acts in favor of both large stored kinetic energy and more 
effective dissipation into heat due to higher probability of collisions.
The partitioning of the injected energy into heat and kinetic motions is 
governed by the ratio of the infall velocity to the local sound velocity.
Only if this ratio is sufficiently small, perturbations in the velocity field 
remain small and can be dissipated fast enough.

\begin{figure}[t]
	\centering
	\resizebox{\hsize}{!}{\includegraphics{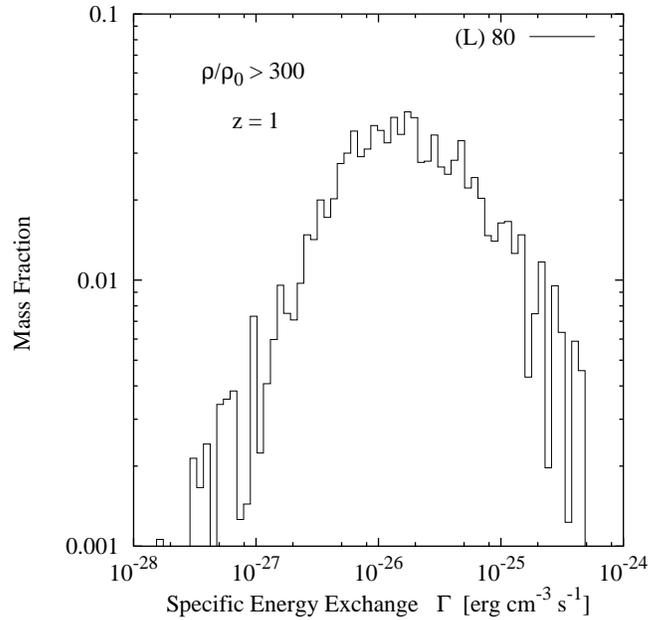}}
	\caption{
	Distribution of the specific energy exchange into regions 
	with density $\rho / \rho_0 > 300$.	
	\label{fig-dsg-dist}
	}
\end{figure}

The specific energy exchange provides the largest possible amount of energy
input into the baryonic matter averaged over galactic scale due to matter
infall, merging, etc..  Figure~\ref{fig-dsg-dist} shows that the expectation
value of the specific energy transfer rate $\Gamma$ amounts to $\approx
10^{-26}\Eerg\Epers\Eperccm$ when regions with density $\rho / \rho_0 > 300$
are considered.  The expectation value is proportional to the density, cf.
Fig.~\ref{fig-spec-2dh}.
The density contrast within the galactic halo is at
least two orders above 300, leading to a transfer rate about 
$10^{-24}\Eerg\Epers\Eperccm$.

We want to estimate whether this energy injection could play a major 
role in the energy balance of the entire baryon content within a galaxy or not. 
The calculated energy injection rate is mainly caused by accretion 
and merging and to some extend by tidal action of neighbored halos. 
We wish to compare this released energy with the supernova (SN) feedback 
averaged over the same considered volume.

From the above obtained specific energy rate the total energy injection rate can 
be obtained for a given volume containing the mass of a galaxy. Indeed, for a 
galaxy mass comparable with 
that of the Milky Way, i.e. of about $10^{10}\Msol$, the total energy injection 
into the corresponding galaxy volume can be estimated by 
$V_{\rm{galax}} \, \Gamma \, n_{\rm{H}} 
	\approx 
	V_{\rm{galax}} \, \Gamma \, ( M_{\rm{galax}}/m_{\rm{H}}/V_{\rm{galax}}) 
	= 
	\Gamma \, M_{\rm{galax}}/m_{\rm{H}}
	\approx 
	4\times 10^{50}\Eerg\Eperyr$.   
The mechanical energy feedback per supernova is of the order of 
$\epsilon\times 10^{51}\Eerg$. The efficiency parameter $\epsilon$ is highly 
uncertain but surely below 0.5; most probably $\epsilon$ is 
even much less \citep[cf.][]{sellwood:99}. Assuming a SN rate of $0.05\Eperyr$ 
\citep[][found for the Milky Way a SN rate of $0.034\pm 
0.028\Eperyr$]{timmes:97} 
we get a mean energy injection rate of $\epsilon\times 10^{49}\Eerg\Eperyr$. 
 
The above provided estimates lead to the conclusion that the energy input rate 
from large-scale evolution uniformly distributed within a 
halo volume comparable with galactic size is certainly comparable at least with 
the energy input rate by SN.

\section{Another prescription for star formation}

\label{sec-sfr}

\begin{figure}[t]
	\centering
	\resizebox{\hsize}{!}{\includegraphics{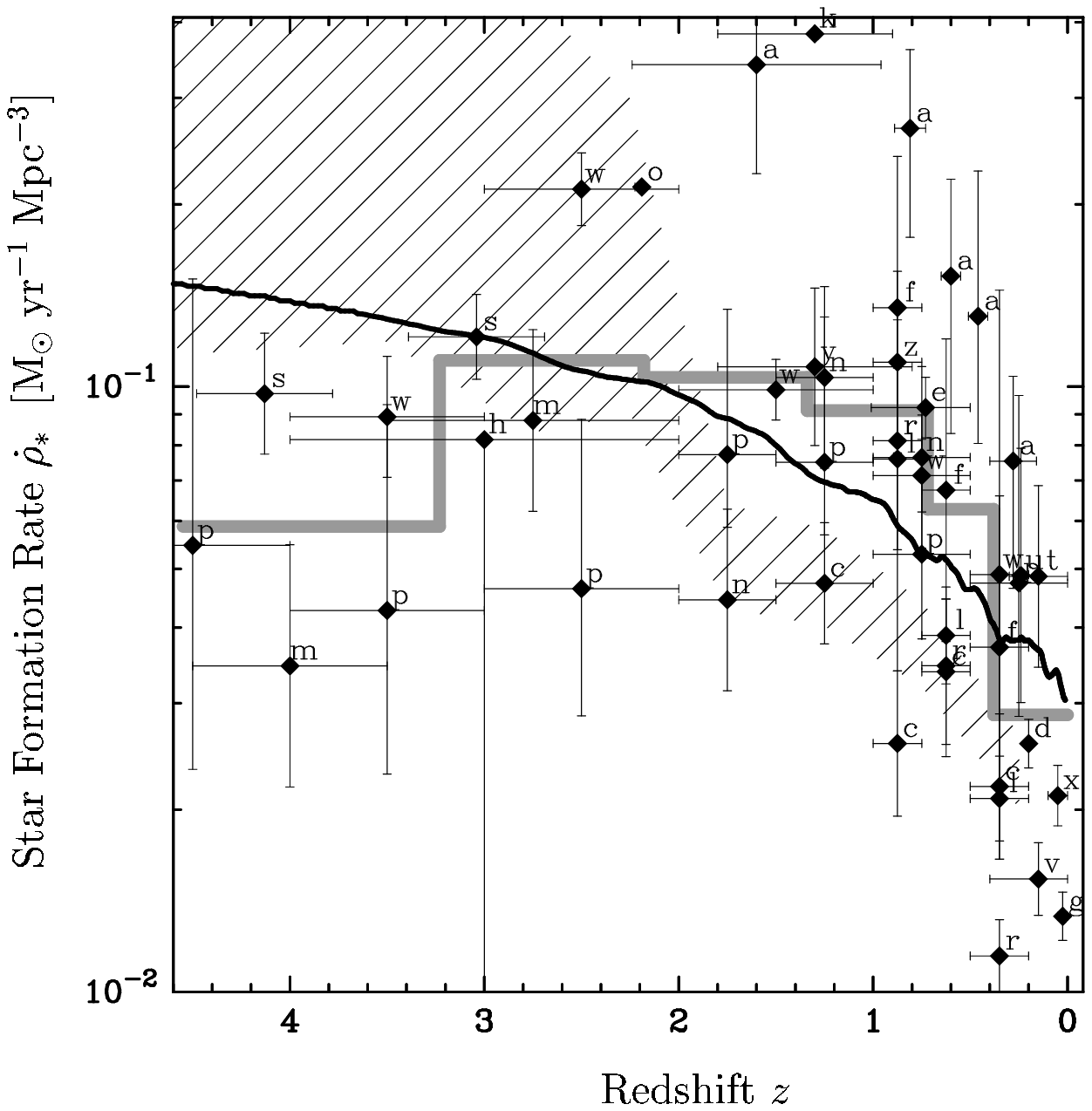}}
	\caption{ 
	The global star formation rate according to 
	Eq.~(\ref{eq-schmidt}),
	taken from simulations with box size $L_{\rm{box}} = 80\,h^{-1}\EMpc$ 
(solid line).
	Depicted data points \citep[cf. the compilation by][]{ascasibar:01} 
	 are from:
	\citet[][p]{pascarelle:98},
	\citet[][m]{steidel:99},
	\citet[][t]{treyer:98},
	\citet[][l]{lilly:96},
	\citet[][c]{cowie:99},
	\citet[][n]{connolly:97},
	\citet[][w]{sawicki:97},
	\citet[][g]{gallego:95},
	\citet[][d]{tresse:98},
	\citet[][g]{glazebrook:99},
	\citet[][y]{yan:99c},
	\citet[][f]{flores:99},
	\citet[][h]{hughes:98},			 
	\citet[][u]{pascual:02},
	\citet[][e]{tresse:02},
	\citet[][k]{hopkins:00},
	\citet[][v]{sullivan:00},
	\citet[][x]{gronwall:99},
	\citet[][o]{moorwood:00},
	\citet[][r]{hammer:97}, and
	\citet[][a]{haarsma:00}.
	We applied the correction given by \citet{ascasibar:01} to obtain star 
formation
   rates in the $\Lambda$CDM model. 	
	The solid gray line indicates the mean of the observed star formation 
rate.
	A quite different evolution is indicated by the data points given by 
\citet{lanzetta:01} (hashed area).	
	}
	\label{fig-sfr}
\end{figure}

For an appropriate description of the IGM in cosmological simulations the 
feedback of stars must be necessarily taken into account. Therefore, a 
prescription for the star formation rate per volume within simulations of the 
large-scale structure formation is needed. 

As briefly reviewed in the introduction it is usually assumed that baryonic 
matter which fulfills certain density and temperature criteria is able to form 
stars according to the Schmidt-law being aware that this is an empirical law, 
based on much different conditions.

The `true' conjunction between the available parameters in such simulations and
star formation is still unknown and allows to speculate and to set up very
different models.  From detailed star formation models in turbulent MCs
indications are provided that the star formation rate may be controlled by
mechanisms which force interstellar turbulence.  One possibility to stir the ISM
may be the merging of galaxies or the occurrence of minor-mergers.  Such merging
events are supposed to generate efficient star formation
\citep[e.g.][]{kauffmann:01}.  

We have investigated the energy-injection into
the internal motions of the halos at sizes of $\approx$ 30 kpc - 2 Mpc due to
the large-scale structure formation and we have found that the energy per mass
input is comparable with energy injection rates attributed to
supernova.  However, even if the energy transfer from the large-scale structure
formation provides sufficient energy to stir the ISM, at present it cannot be
shown neither a mechanism to propagate the injected energy down to scales of
observed ISM turbulence nor the detailed structure of the necessary turbulent
field \citep[for addressing this problem see however][]{jog:88,sellwood:99}.
Although the effect of the energy transfer onto the ISM is
unknown, it represents one of the possible sources for driving the interstellar 
turbulence. Therefore, we assume here that the energy input is the linking 
parameter and make the following proposition:

The energy transfer from the large-scale structure formation provides sufficient
energy to stir the ISM and the transfer is the main energy source for
interstellar turbulence.  Then, as a consequence, the amount of star formation 
should be
determined by this energy transfer as well.

Following this quite speculative conjunction we would like to construct a
prescription for the star formation rate within a small volume still described
within large-scale structure formation simulations, i.e. related to scales of a
few kpc.  We assume direct proportionality to the local energy injection rate
$\pi_K^>(\vx_i)\propto \partial_t \sum_{n \in i} \sigma_n^2 $.
For a given halo part of the kinetic energy of the gas is transformed into 
heat. High temperature is expected to suppress star formation. We make the crude 
assumption that the thermal energy is a constant fraction proportional to the 
kinetic energy of the halo gas. This leads us to assume inverse
proportionality of the SFR to the energy already stored within the small-scale 
motions $e_K^>(\vx_i) \propto \sum_{n \in i} \sigma_n^2 $. 

In the result we get for the fraction of mass which is transferred into stars 
during a certain time interval $\Delta t$ within a given cell $\vx_i$ 
\beqa
\frac{ \Delta m_\ast	(\vx_i)}{ m (\vx_i) }
	&=&
	\epsilon_0 \;
	\frac{\pi_K^>(\vx_i)}{e_K^>(\vx_i)}  \Delta t
	\nn
	\\
	& \propto &
	\frac{\partial_t  \sum_{n \in i} \sigma_n^2 }
		{\sum_{n \in i} \sigma_n^2 }	 \Delta t
	=
	\frac{ \partial_t \langle \sigma_n^2 \rangle_{\rm{cell}} }
		  {          	\langle \sigma_n^2 \rangle_{\rm{cell}} }
 	\Delta t
	\label{eq-schmidt}
\eeqa	
where $\epsilon_0$ is a normalization constant which has to be determined, e.g. 
by 
the observed SFR at any fixed redshift, and  $m (\vx_i)$ denotes the mass within 
the considered cell.
As can be seen from the last term in Eq.~(\ref{eq-schmidt}) the ratio $\pi_K^>/e_K^>$ is 
independent from the local density.
The comoving cosmic SFR $\dot{\rho}_\ast$ is obtained by summing up the 
contributions 
of all cells and dividing by the comoving volume of the simulation box 
$L_{\rm{box}}^3$
\beqa
\dot{\rho}_\ast
	&=&
	\frac{\sum_i \, \Delta m_\ast	(\vx_i)}{\Delta t \; L_{\rm{box}}^3}
	.
	\label{eq-cosmic-sfr}
\eeqa	
Note, assuming for the SFR proportionality to the local specific energy 
injection 
the expression (\ref{eq-schmidt}) represents the simplest 
combination of quantities obeying the dimensional requirements.

Since the energy injection rate is closely correlated with the merging and the 
matter infall rates as was shown in an earlier section the star formation 
history calculated in this manner
reflects mainly the influence of merging processes. 

\begin{figure}[t]
	\centering
	\resizebox{\hsize}{!}{\includegraphics{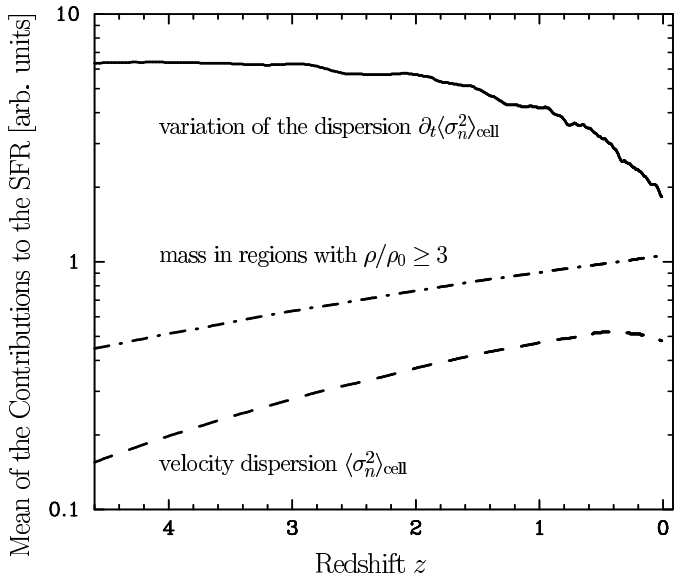}}
	\caption{ 
	Evolution of the mean of the different constituents 
	in the prescription Eq.~(\ref{eq-schmidt}),
	namely the mean of the variation of the local velocity dispersion 
	$\partial_t \langle \sigma_n^2 \rangle_{\rm{cell}}$, 
	the amount of mass contained in regions with density contrast $\rho/\rho_0 \ge 3$
	(cf. Sec~\ref{sec-e-ex}),
	and the mean of the local velocity dispersion
	$\langle \sigma_n^2 \rangle_{\rm{cell}}$.
	}
	\label{fig-sfr-sep}
\end{figure}

For transparency we give the time evolution of the different constituents 
of the expression (\ref{eq-schmidt}) averaged over the comoving box volume,
see Fig.~\ref{fig-sfr-sep}. 
The amount of mass 
concentrated in regions with density contrast $\rho / \rho_0 \ge 3$  
increases continuously with time. The hierarchical clustering 
process leads to increasing kinetic energy of the halo matter which happens 
according to the mass inflow and concentration. This explains the quite similar 
slope for the density and kinetic energy evolution. The energy injection into 
small scales is a slightly decreasing function with time as shown already in the 
preceding section, cf. Fig.~\ref{fig-spec}.

Implementing the prescription Eq.~(\ref{eq-schmidt}) 
within our simulations 
we obtain according to Eq.~(\ref{eq-cosmic-sfr})
the evolution of the cosmic star formation rate $\dot{\rho}_\ast$, 
as shown in Fig.~\ref{fig-sfr}. 
For comparison we included a compilation of observational data of  
star formation rates recently obtained for different redshifts. 
We have fixed the normalization constant $\epsilon_0$ using the data in the 
vicinity of $z = 0.5$.
Even if there is a big scatter in the estimated star formation rates it is 
commonly accepted that there is a steep increase from redshift zero to redshift 
one (cf. gray line).
Estimates at higher redshift indicate a subsequent decline even if these 
measurements are still under debate.
Although dust extinction is to some extend allowed for it may still cause the 
decline of the data points.
A quite different star formation history is recently given by 
\citet{lanzetta:01}. 
They found a continuous increase of the SFR with redshift.

The evolution of the calculated 'star formation rate' matches the cosmic star 
formation history to some extent.
However, the limitations of the proposed model are obvious:
In addition to the considered impact of energy transfer, which mainly reflects 
the impact of merging and accretion, the star formation is 
affected by hydrodynamical effects, radiation,  etc. related to the 
properties of the baryonic matter. 
In particular, the back-reaction of the already formed stars onto the gas 
by winds, supernova etc. may locally turn off any subsequent star formation.  
These processes result almost exclusively in an reduction of star 
formation which may cause in particular a steeper slope at small redshifts than 
obtained here. 

The advantage of the proposed quantity is obvious as well:  It can be computed
on-the-flight without determining the single matter halos and without tracing 
back
their evolution.  It can be used complementary within the framework of
semi-analytical descriptions of star formation applied to the results of n-body
simulations.  Of course, this cannot replace the knowledge of the detailed star
formation process including the full cooling history and detailed velocity
fields.  However, for the consideration of large-scale structure formation and
the large scale distribution of quantities related to star formation this could
serve as a suitable approximation.

\section{Summary}\label{summary}

During the formation of the large-scale structures gravitational energy is 
transferred into the internal motion of the dark matter halos.
In particular by merging of the halos part of the kinetic energy is subsequently 
transferred from the large-scale movement into small-scale internal motions.
The baryonic matter possesses the same amount of kinetic energy per mass as long 
as it moves together with the dark matter, i.e. as long as gravitational 
interaction dominates over pressure forces.
Thus the large-scale structure formation is transferring energy into small 
scales. We have addressed the question how this transfer can be described, how 
it evolves during the cosmological structure formation and whether this 
transferred energy rate is of comparable order of magnitude to compete with 
internal energy sources in the galaxies.  

We have performed cosmological n-body simulations to determine which energy 
amount the large-scale structure formation provides to a given small scale.

A crucial point in that picture is how the energy transfer is directed.
We have considered this question investigating the evolution of the power 
spectrum of the large-scale velocity field.
To this end we have used a method introduced by \citet{lomb:76}. 
The power spectrum evolves according to the linear growth at scales larger 
$\approx 60\,h^{-1}\EMpc$.
Below these scales it possesses a more shallow region and at scales below 
$\approx 0.5\,h^{-1}\EMpc$ it roughly behaves like a power law $\sim k^{-4.2}$.
In general, the features of the velocity spectrum are similar to those of the 
power spectrum of the density field.
The behavior of the velocity spectra shows that large-scale modes dominate the 
power of small-scale modes and that the energy transfer throughout the modes is 
almost exclusively one-directional, namely from larger to smaller scales.

To calculate the energy transfer through a given scale we have used the concept 
of cumulative energy.
The Fourier transforms of the velocity field and of the density of momentum are 
subdivided into a high-pass and a low-pass filtered part.
As a result the energy density can also be decomposed into a high-pass and a 
low-pass filtered part and the scale-by-scale energy budget contains a term 
describing the energy exchange between the two spectral ranges.
The exchange rate may be calculated by help of the time derivative of the 
high-pass filtered energy and equals to the energy injection rate into scales 
smaller than the given cut-off length.

Using the involved quantities we can define straightforwardly the mean global 
energy exchange rate into given scales at any time during the performed n-body 
simulations. 
We obtain the mean energy exchange rate which increases with redshift according 
to a power law $\sim (z+1)^{3}$.  
We argue that this can be attributed to the mass increase of dark matter halos. 
The evolution includes all kinds of infall, i.e.  continuous accretion of 
matter onto halos, minor mergers, or even merging of halos. 
If all processes evolve similar or if merging is dominant the evolution of the 
mean energy exchange exhibits the same time-dependence as the merging rate of 
dark matter halos.

Due to the additivity of the cumulative energy with respect to its local 
contributions if applied to grid based n-body systems a {\em local} high-pass 
filtered mean energy exchange rate can be 
defined supposing that the size of the local averaging volume is equal to the 
cut-off scale $\lambda_K$.
A specific energy exchange rate is given by the ratio of the local energy 
exchange rate into a given volume to the local mass density.
Using this definition we obtain the mean specific energy exchange rate which is 
as
low as $\approx 10^{-28}\,n\Eerg\Epers\Eperccm$.
However, its expectation value is proportional to the density. 
Thus, in regions as dense as the galactic halo the energy input due to the 
energy exchange may be as large as $\approx 10^{-24}\,n\Eerg\Epers\Eperccm$,
which is comparable with the kinetic energy feedback by supernova averaged 
over the same galactic volume.

On the basis of our results we consider a speculative picture: 
We assume  that the injected energy propagates to MC scales and is 
therefore available for driving the MC turbulence leading to star formation.
On the ground of this quite heuristic assumptions we propose an estimator for 
the star formation rate in a local volume in cosmological simulations:
$\dot{m}_\ast \propto (\pi_K^> / e_K^>) \: m$.

The model reproduces approximately the observed global evolution of the star 
formation rate.
Adopting this model means that merging and matter 
infall processes are considered to enhance the star formation at least or  
to be the leading processes.

\newcommand{\nucphys}{Nucl. Phys.}
\newcommand{\PRL}{Phys. Rev. Lett.}
\newcommand{\TUB}{Technische Universit{\"a}t Berlin}
\newcommand{\ZfPhys}{Z. Phys.}

\bibliography{ne-astro,ne-nuc,ne-books}

\begin{thebibliography}{56}
\expandafter\ifx\csname natexlab\endcsname\relax\def\natexlab#1{#1}\fi

\bibitem[{{Ascasibar} {et~al.}(2001){Ascasibar}, Yepes, Gottl\"ober, \&
  M\"uller}]{ascasibar:01}
{Ascasibar}, Y., Yepes, G., Gottl\"ober, S., \& M\"uller, V. 2001, {A\&A},
  submitted

\bibitem[{{Ballesteros-Paredes} {et~al.}(1999){Ballesteros-Paredes}, {V{\'
  a}zquez-Semadeni}, \& {Scalo}}]{ballesteros-paredes:99}
{Ballesteros-Paredes}, J., {V{\' a}zquez-Semadeni}, E., \& {Scalo}, J. 1999,
  \apj, 515, 286

\bibitem[{{Blitz} {et~al.}(1999){Blitz}, {Spergel}, {Teuben}, {Hartmann}, \&
  {Burton}}]{blitz:99b}
{Blitz}, L., {Spergel}, D.~N., {Teuben}, P.~J., {Hartmann}, D., \& {Burton},
  W.~B. 1999, \apj, 514, 818

\bibitem[{{Blitz} \& {Williams}(1999)}]{blitz:99}
{Blitz}, L. \& {Williams}, J.~P. 1999, in NATO ASIC Proc. 540: The Origin of
  Stars and Planetary Systems, 3+

\bibitem[{{Burkert}(2001)}]{burkert:01}
{Burkert}, A. 2001, in 9 pages, 2 figures, conference proceeding. to appear in
  "Modes of Star Formation", eds. E.K. Grebel and W. Brandner, 5298+

\bibitem[{{Carlberg} {et~al.}(1994){Carlberg}, {Pritchet}, \&
  {Infante}}]{carlberg:94}
{Carlberg}, R.~G., {Pritchet}, C.~J., \& {Infante}, L. 1994, \apj, 435, 540

\bibitem[{{Connolly} {et~al.}(1997){Connolly}, {Szalay}, {Dickinson},
  {Subbarao}, \& {Brunner}}]{connolly:97}
{Connolly}, A.~J., {Szalay}, A.~S., {Dickinson}, M., {Subbarao}, M.~U., \&
  {Brunner}, R.~J. 1997, \apjl, 486, L11

\bibitem[{{Cowie} {et~al.}(1999){Cowie}, {Songaila}, \& {Barger}}]{cowie:99}
{Cowie}, L.~L., {Songaila}, A., \& {Barger}, A.~J. 1999, \aj, 118, 603

\bibitem[{Elmegreen(2000)}]{elmegreen:00}
Elmegreen, B.~G. 2000, ApJ, 530, 277

\bibitem[{{Flores} {et~al.}(1999){Flores}, {Hammer}, {Thuan}, {C{\' e}sarsky},
  {Desert}, {Omont}, {Lilly}, {Eales}, {Crampton}, \& {Le F{\`
  e}vre}}]{flores:99}
{Flores}, H., {Hammer}, F., {Thuan}, T.~X., {C{\' e}sarsky}, C., {Desert},
  F.~X., {Omont}, A., {Lilly}, S.~J., {Eales}, S., {Crampton}, D., \& {Le F{\`
  e}vre}, O. 1999, \apj, 517, 148

\bibitem[{Frisch(1994)}]{frisch:95}
Frisch, U. 1994, Turbulence: The Legacy of A. N. Kolmogorov (Heidelberg:
  Spetrum, Akad. Verlad)

\bibitem[{{Gallego} {et~al.}(1995){Gallego}, {Zamorano}, {Aragon-Salamanca}, \&
  {Rego}}]{gallego:95}
{Gallego}, J., {Zamorano}, J., {Aragon-Salamanca}, A., \& {Rego}, M. 1995,
  \apjl, 455, L1

\bibitem[{{Glazebrook} {et~al.}(1999){Glazebrook}, {Blake}, {Economou},
  {Lilly}, \& {Colless}}]{glazebrook:99}
{Glazebrook}, K., {Blake}, C., {Economou}, F., {Lilly}, S., \& {Colless}, M.
  1999, \mnras, 306, 843

\bibitem[{{Goldman}(2000)}]{goldman:00}
{Goldman}, I. 2000, \apj, 541, 701

\bibitem[{Gottl\"ober {et~al.}(2000)Gottl\"ober, Klypin, \&
  Kravtsov}]{gottloeber:00}
Gottl\"ober, S., Klypin, A., \& Kravtsov, A.~V. 2000, preprint,
  astro-ph/0004132

\bibitem[{{Gronwall}(1999)}]{gronwall:99}
{Gronwall}, C. 1999, in After the Dark Ages: When Galaxies were Young (the
  Universe at 2 < z < 5). 9th Annual October Astrophysics Conference in
  Maryland held 12-14 October, 1998. College Park, Maryland. Edited by S. Holt
  and E. Smith. American Institute of Physics Press, 1999, p. 335, 335+

\bibitem[{{Haarsma} {et~al.}(2000){Haarsma}, {Partridge}, {Windhorst}, \&
  {Richards}}]{haarsma:00}
{Haarsma}, D.~B., {Partridge}, R.~B., {Windhorst}, R.~A., \& {Richards}, E.~A.
  2000, \apj, 544, 641

\bibitem[{{Hammer} {et~al.}(1997){Hammer}, {Flores}, {Lilly}, {Crampton}, {Le
  Fevre}, {Rola}, {Mallen-Ornelas}, {Schade}, \& {Tresse}}]{hammer:97}
{Hammer}, F., {Flores}, H., {Lilly}, S.~J., {Crampton}, D., {Le Fevre}, O.,
  {Rola}, C., {Mallen-Ornelas}, G., {Schade}, D., \& {Tresse}, L. 1997, \apj,
  481, 49+

\bibitem[{{Hopkins} {et~al.}(2000){Hopkins}, {Connolly}, \&
  {Szalay}}]{hopkins:00}
{Hopkins}, A.~M., {Connolly}, A.~J., \& {Szalay}, A.~S. 2000, \aj, 120, 2843

\bibitem[{{Hughes} {et~al.}(1998){Hughes}, {Serjeant}, {Dunlop},
  {Rowan-Robinson}, {Blain}, {Mann}, {Ivison}, {Peacock}, {Efstathiou}, {Gear},
  {Oliver}, {Lawrence}, {Longair}, {Goldschmidt}, \& {Jenness}}]{hughes:98}
{Hughes}, D.~H., {Serjeant}, S., {Dunlop}, J., {Rowan-Robinson}, M., {Blain},
  A., {Mann}, R.~G., {Ivison}, R., {Peacock}, J., {Efstathiou}, A., {Gear}, W.,
  {Oliver}, S., {Lawrence}, A., {Longair}, M., {Goldschmidt}, P., \& {Jenness},
  T. 1998, \nat, 394, 241

\bibitem[{{Jog} \& {Ostriker}(1988)}]{jog:88}
{Jog}, C.~J. \& {Ostriker}, J.~P. 1988, \apj, 328, 404

\bibitem[{Kauffmann {et~al.}(2001)Kauffmann, Charlot, \& Balogh}]{kauffmann:01}
Kauffmann, G., Charlot, S., \& Balogh, M.~L. 2001, preprint, astro-ph/0103130

\bibitem[{{Kauffmann} {et~al.}(1999){Kauffmann}, {Colberg}, {Diaferio}, \&
  {White}}]{kauffmann:99}
{Kauffmann}, G., {Colberg}, J. .~M., {Diaferio}, A., \& {White}, S. D.~M. 1999,
  MNRAS, 307, 529

\bibitem[{{Kay} {et~al.}(2001){Kay}, {Pearce}, {Frenk}, \& {Jenkins}}]{kay:01}
{Kay}, S.~T., {Pearce}, F.~R., {Frenk}, C.~S., \& {Jenkins}, A. 2001, accepted
  for publication in MNRAS, astro-ph/0106462

\bibitem[{{Kennicutt}(1998)}]{kennicutt:98}
{Kennicutt}, R.~C. 1998, \apj, 498, 541+

\bibitem[{Klessen(2000)}]{klessen:00b}
Klessen, R.~S. 2000, ASP Conference Series, T. Montmerle \& Ph. Andre, eds.,
  preprint, astro-ph/0011224

\bibitem[{{Klypin} {et~al.}(2000){Klypin}, {Kravtsov}, \&
  {Col{\'i}n}}]{klypin:99}
{Klypin}, A., {Kravtsov}, A., \& {Col{\'i}n}, P. 2000, in ASP Conf. Ser. 201:
  Cosmic Flows Workshop, 344+

\bibitem[{{Kolatt} {et~al.}(1999){Kolatt}, {Bullock}, {Somerville}, {Sigad},
  {Jonsson}, {Kravtsov}, {Klypin}, {Primack}, {Faber}, \& {Dekel}}]{kolatt:99}
{Kolatt}, T.~S., {Bullock}, J.~S., {Somerville}, R.~S., {Sigad}, Y., {Jonsson},
  P., {Kravtsov}, A.~V., {Klypin}, A.~A., {Primack}, J.~R., {Faber}, S.~M., \&
  {Dekel}, A. 1999, \apjl, 523, L109

\bibitem[{{Lanzetta} {et~al.}(2001){Lanzetta}, {Yahata}, {Pascarelle}, {Chen},
  \& {Fernandez-Soto}}]{lanzetta:01}
{Lanzetta}, K.~M., {Yahata}, N., {Pascarelle}, S., {Chen}, H., \&
  {Fernandez-Soto}, A. 2001, in 28 pages, 9 figures; accepted for publication
  in the Astrophysical Journal., 11129+

\bibitem[{{Larson}(2001)}]{larson:01}
{Larson}, R.~B. 2001, in Summary talk at the meeting on "Modes of Star
  Formation and the Origin of Field Populations", Heidelberg, Germany, October
  2000; to be published in the ASP Conference Series, edited by E. K. Grebel
  and W. Brandner, 1046+

\bibitem[{{Lilly} {et~al.}(1996){Lilly}, {Le Fevre}, {Hammer}, \&
  {Crampton}}]{lilly:96}
{Lilly}, S.~J., {Le Fevre}, O., {Hammer}, F., \& {Crampton}, D. 1996, \apjl,
  460, L1

\bibitem[{{Lomb}(1976)}]{lomb:76}
{Lomb}, N.~R. 1976, \apss, 39, 447

\bibitem[{{Mac Low} {et~al.}(1998){Mac Low}, {Klessen}, {Burkert}, \&
  {Smith}}]{maclow:98}
{Mac Low}, M.-M., {Klessen}, R.~S., {Burkert}, A., \& {Smith}, M.~D. 1998,
  Phys. Rev. Lett., 80, 2754

\bibitem[{{Moorwood} {et~al.}(2000){Moorwood}, {van der Werf}, {Cuby}, \&
  {Oliva}}]{moorwood:00}
{Moorwood}, A.~F.~M., {van der Werf}, P.~P., {Cuby}, J.~G., \& {Oliva}, E.
  2000, \aap, 362, 9

\bibitem[{{Nagamine} {et~al.}(2001){Nagamine}, {Fukugita}, {Cen}, \&
  {Ostriker}}]{nagamine:01}
{Nagamine}, K., {Fukugita}, M., {Cen}, R., \& {Ostriker}, J.~P. 2001, \apj,
  558, 497

\bibitem[{{Obukhov}(1941)}]{obukhov:41}
{Obukhov}, A.~M. 1941, Izv. Akad. Nauk SSSR Ser. Geogr. Geofiz., 5, 453

\bibitem[{{Padoan} {et~al.}(2001){Padoan}, {Juvela}, {Goodman}, \&
  {Nordlund}}]{padoan:01}
{Padoan}, P., {Juvela}, M., {Goodman}, A.~A., \& {Nordlund}, {\AA}. 2001, \apj,
  553, 227

\bibitem[{{Pascarelle} {et~al.}(1998){Pascarelle}, {Lanzetta}, \& {Fern{\'
  a}ndez-Soto}}]{pascarelle:98}
{Pascarelle}, S.~M., {Lanzetta}, K.~M., \& {Fern{\' a}ndez-Soto}, A. 1998,
  \apjl, 508, L1

\bibitem[{{Pascual} {et~al.}(2002){Pascual}, {Gallego}, {Aragon-Salamanca}, \&
  {Zamorano}}]{pascual:02}
{Pascual}, S., {Gallego}, J., {Aragon-Salamanca}, A., \& {Zamorano}, J. 2002,
  mNRAS, in press (astro-ph/0110177)

\bibitem[{{Press} {et~al.}(1986){Press}, {Flannery}, \& {Teukolsky}}]{press:86}
{Press}, W.~H., {Flannery}, B.~P., \& {Teukolsky}, S.~A. 1986, Numerical
  recipes. The art of scientific computing (Cambridge: University Press, 1986)

\bibitem[{{Sawicki} {et~al.}(1997){Sawicki}, {Lin}, \& {Yee}}]{sawicki:97}
{Sawicki}, M.~J., {Lin}, H., \& {Yee}, H.~K.~C. 1997, \aj, 113, 1

\bibitem[{{Scalo} \& {Biswas}(2001)}]{scalo:02}
{Scalo}, J. \& {Biswas}, A. 2001, \mnras, accepted, (astro-ph/0111201)

\bibitem[{{Scargle}(1982)}]{scargle:82}
{Scargle}, J.~D. 1982, \apj, 263, 835

\bibitem[{{Schmidt}(1959)}]{schmidt:59}
{Schmidt}, M. 1959, \apj, 129, 243+

\bibitem[{{Sellwood} \& {Balbus}(1999)}]{sellwood:99}
{Sellwood}, J.~A. \& {Balbus}, S.~A. 1999, \apj, 511, 660

\bibitem[{{Somerville} {et~al.}(2001){Somerville}, {Primack}, \&
  {Faber}}]{somerville:01}
{Somerville}, R.~S., {Primack}, J.~R., \& {Faber}, S.~M. 2001, \mnras, 320,
  504+

\bibitem[{{Springel}(2000)}]{springel:00}
{Springel}, V. 2000, MNRAS, 312, 859

\bibitem[{{Steidel} {et~al.}(1999){Steidel}, {Adelberger}, {Giavalisco},
  {Dickinson}, \& {Pettini}}]{steidel:99}
{Steidel}, C.~C., {Adelberger}, K.~L., {Giavalisco}, M., {Dickinson}, M., \&
  {Pettini}, M. 1999, \apj, 519, 1

\bibitem[{{Steinmetz}(2001)}]{steinmetz:01}
{Steinmetz}, M. 2001, in ASP Conf. Ser. 230: Galaxy Disks and Disk Galaxies,
  633--640

\bibitem[{{Sullivan} {et~al.}(2000){Sullivan}, {Treyer}, {Ellis}, {Bridges},
  {Milliard}, \& {Donas}}]{sullivan:00}
{Sullivan}, M., {Treyer}, M.~A., {Ellis}, R.~S., {Bridges}, T.~J., {Milliard},
  B., \& {Donas}, J.~. 2000, \mnras, 312, 442

\bibitem[{{Timmes} {et~al.}(1997){Timmes}, {Diehl}, \& {Hartmann}}]{timmes:97}
{Timmes}, F.~X., {Diehl}, R., \& {Hartmann}, D.~H. 1997, \apj, 479, 760+

\bibitem[{{Tresse} \& {Maddox}(1998)}]{tresse:98}
{Tresse}, L. \& {Maddox}, S.~J. 1998, \apj, 495, 691+

\bibitem[{{Tresse} {et~al.}(2002){Tresse}, {Maddox}, {Le Fevre}, \&
  {Cuby}}]{tresse:02}
{Tresse}, L., {Maddox}, S.~J., {Le Fevre}, O., \& {Cuby}, J.-G. 2002, mNRAS,
  submitted (astro-ph/0111390)

\bibitem[{{Treyer} {et~al.}(1998){Treyer}, {Ellis}, {Milliard}, {Donas}, \&
  {Bridges}}]{treyer:98}
{Treyer}, M.~A., {Ellis}, R.~S., {Milliard}, B., {Donas}, J., \& {Bridges},
  T.~J. 1998, \mnras, 300, 303

\bibitem[{{Yan} {et~al.}(1999){Yan}, {McCarthy}, {Freudling}, {Teplitz},
  {Malumuth}, {Weymann}, \& {Malkan}}]{yan:99c}
{Yan}, L., {McCarthy}, P.~J., {Freudling}, W., {Teplitz}, H.~I., {Malumuth},
  E.~M., {Weymann}, R.~J., \& {Malkan}, M.~A. 1999, \apjl, 519, L47

\bibitem[{{Yepes} {et~al.}(1997){Yepes}, {Kates}, {Khokhlov}, \&
  {Klypin}}]{yepes:97}
{Yepes}, G., {Kates}, R., {Khokhlov}, A., \& {Klypin}, A. 1997, \mnras, 284,
  235

\end{thebibliography}
\bibliographystyle{apj}

\end{document}